\documentclass{llncs}

\usepackage{graphicx}

\usepackage[latin1]{inputenc}

\begin{document}

\mainmatter              % start of the contributions
\title{An Ant-Based Model for Multiple Sequence Alignment}
\titlerunning{An Ant-Based Model for Multiple Sequence Alignment}  
\author{Fr\'ed\'eric Guinand \and Yoann Pign\'e\thanks{Authors are
    alphabetically sorted. The work of Y. Pign\'e is partially
    supported by French Ministry of Research and Higher Education.}}

\authorrunning{Fr\'ed\'eric Guinand \and Yoann Pign\'e}   % abbreviated author list (for running head)
\tocauthor{Fr\'ed\'eric Guinand, Yoann Pign\'e}
\institute{LITIS laboratory, Le Havre University, France\\
\texttt{www.litislab.eu}}

\maketitle              % typeset the title of the contribution

\vspace{-0.3cm}

\begin{abstract}
Multiple sequence alignment is a key process in today's biology, and finding a
relevant alignment of several sequences is much more challenging than
just optimizing some improbable evaluation functions. 
Our approach for addressing multiple sequence alignment focuses on the
building of structures in a new graph model: the factor graph model.
This model relies on block-based formulation of the original problem, 
formulation that seems to be one of the most suitable ways for capturing
evolutionary aspects of alignment.
The structures are implicitly built by a colony of ants laying down
pheromones in the factor graphs, according to relations between blocks
belonging to the different sequences.
\end{abstract}

\vspace{-0.5cm}

%-------------------------------------------------------------------------------
\section{Introduction}
\label{sec:intro}
%-------------------------------------------------------------------------------

For years, manipulation and study of biological sequences have been added to
the set of common tasks performed by biologists in their daily activities.
Among the numerous analysis methods, multiple sequence alignment (MSA) is probably
one of the most used. 
%A multiple sequence alignment (MSA) is an alignment of some nucleotide
%or amino acid-based sequences. 
Biological sequences come from actual living beings, and the role of MSA
consists in exhibiting the similarities and differences 
between them. Considering sets of homologous sequences, differences
may be used to assess the evolutionary distance between species in the context
of phylogeny. The results of this analysis may
also be used to determine conservation of protein domains or structures. 
%But MSA suffers from a lack of a biologically relevant evaluation function. 
While most of the time the process is performed for aligning a limited number
of thousands bp-long sequences, it can also be used at the genome level
allowing biologists to discover 
new features that could not be exhibited at a lower level of study
\cite{mummer}. 
% Lagan is a tool for Whole genome alignment which implements a glocal alignment
%algorithm which first finds regions of local similarity before applying a
%global alignment across those seeds.  
%Main tools for performing genome-wide comparison are Mummer, Waba, rely on 
In all cases, one of the major difficulties is the determination of a biologically relevant
alignment, performed without relying explicitly on evolutionary information
like a phylogenetic tree.

%\begin{figure}[!ht]
%  \begin{center}
%  \includegraphics[scale=0.6]{images/short_blocks_sace.eps}
%  \end{center}
%  \caption{A view of genes for protein coding in Chromosom I of {\em Saccharomyces
%    cerevisiae}. Factors may be identified to these substrings obtained from
%    biological analyses. \label{fig:block}}
%\end{figure}

Among existing approaches for determining such relevant alignments,
one of them rests on the notion of block. 
A block is a set of factors present in several sequences.
Each factor belonging to one block is an almost identical substring. 
It may correspond to a highly conserved zone from an evolutionary point of
view.  
Starting from the set of factors for each sequence, the problem we address
is the building of blocks.
It consists in choosing and gathering almost identical factors common to
several sequences in the most appropriate way, given that one block cannot
contain more than one factor per sequence, that each factor can belong
to only one block and that two blocks cannot cross each other.
For building such blocks, we propose an approach based on ant
colonies. This problem is very close to some classical optimization issues
except that the process does not use any evaluation function since it seems
unlikely to find a biologically relevant one.
As such, it also differs notably from other works in the domain setting up ant
colonies for computing alignments for a set of biological sequences 
\cite{antmsa2003,antmsa2006}.

The following section details the proposed graph model. Sect. \ref{sec:algorithm} goes deeper into the ant algorithm details. Finally, Sect. \ref{sec:analysis} studies the behavior of the algorithm with examples.

\vspace{-0.2cm} 

%-------------------------------------------------------------------------------
\section{Model}
\label{sec:model}
%-------------------------------------------------------------------------------

There exist many different families of algorithms for determining multiple
sequence alignments, dynamic programming, progressive or iterative methods,
motif-based approaches... However, if the number of methods is
important, the number of models on which these methods operate is much more
limited.  
Indeed, most algorithms use to consider nucleotide sequences either
as strings or as graphs. In any case however, the problem is formulated as an
optimization problem and an evaluation function is given. 
Within this paper, we propose another approach based on a graph of factors,
where the factors are sub-sequences present in, at least, two sequences. 
Instead of considering these factors individually, the formulation considers
that they interact with each other when they are neighbors in different
sequences, such that our {\em factor graph} may be understood as a factor/pattern
interaction network.
Considering such a graph, a multiple sequence alignment corresponds to a set
of structures representing highly interacting sets of factors. 
The original goal may be now expressed as the detection of such structures and
we propose to perform such a task with the help of artificial ants.

%--------------------------------------------  
\subsection{Graph Model}
\label{ssec:graphModel}
%--------------------------------------------

\begin{figure}
\centering 
\includegraphics[width=0.7\textwidth]{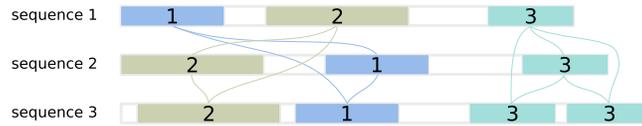}
\caption{This is a set of three sequences. Common subsequences
  of these sequences which are repeated are labeled. After the conversion,
  each sequence of the alignment is displayed as a list of factors. Here
  sequence 1 = [1,2,3], sequence 2 = [2,1,3] and sequence 3 = [2,1,3,3]. Thin
  lines link factors that may be aligned together.} 
\label{fig:sequences}
\end{figure}

An alignment is usually displayed sequence by sequence, with the nucleotides 
or amino acids that compose it. 
Here the interest is given to the factors that
compose each sequence. So each sequence of the alignment is displayed as a
list of the factors it is composed of. Fig. \ref{fig:sequences} illustrates
such a representation, where sequences are displayed as series of factors.  

%\begin{figure}
%\centering 
%\includegraphics[width=0.7\textwidth]{images/alignment.eps}
%\caption{One possible block alignment from the set of sequences shown Fig. \ref{fig:sequences}. Each column in the alignment is a block. The block made of ``2'' factors is a \emph{complete} one. }
%\label{fig:alignment} 
%\end{figure}

There exists a relation between factors (named 1, 2 and 3 in Fig.
\ref{fig:sequences}) as soon as there are almost identical. 
Indeed, two identical factors on different sequences may be aligned. 
Such an alignment aims at creating blocks. Together with factors, these
relations can be represented by a graph $G=(V,E)$.
The set $V$ of nodes represents all the factors appearing in the sequences, and
edges of $E$ link factors that may be aligned. 
These graphs are called \emph{factor graphs}. 
A \emph{factor graph} is a complete graph where edges
linking factors attending on the same sequence are removed. Indeed,
a given factor $f$ may align with any other identical factor $f'$ provided
$f'$ does not belong to the same sequence. In
Fig. \ref{fig:sequences}, thin links between factors illustrate the possible
alignments between them. 
%Note on the sequence 3 that no link exists between the two factors named "3" since they cannot be align
%together. 

From a graph point of view the sequential order "sequence by sequence" has no
sense. The alignment problem is modeled as a set of \emph{factor graphs}. So as to
differentiate the different factors, they are given a unique identifier. 
Each factor is assigned a triplet $[x,y,z]$ where $x$ is an
identifier for the pattern, $y$ is the identifier of the sequence the factor
is located on and $z$ is the occurrence of this pattern on the given
sequence. For instance, on Fig. \ref{fig:sequences}, the bottom right factor
of the third sequence is identified by the triplet $[3,3,2]$. Namely, it is the
pattern "3", located on the sequence 3 and it occurs for the second time on
this sequence, given the sequences are read from left to right. 

\begin{figure}
        \centering
                \includegraphics[width=0.80\textwidth]{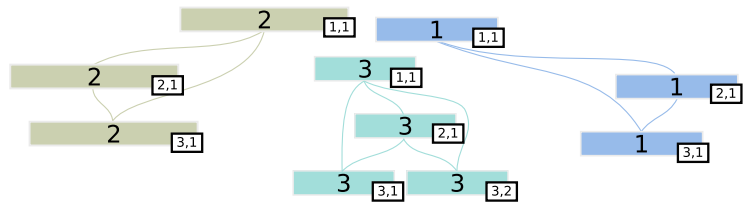}
        \caption{The alignment seen in  Fig. \ref{fig:sequences} displayed as a set of \emph{factor graphs}.}
        \vspace{-0.5cm} 

         \label{fig:graph}       
\end{figure}

Using that model the sequences are not ordered as it is the case when considering
progressive alignment methods.
Fig. \ref{fig:graph} illustrates such graphs according to the representation of
Fig. \ref{fig:sequences}.  

From each \emph{factor graph} a subset of factors may be selected to create a
block. If the block is composed of one factor per sequence, it is a
complete block, but if one sequence is missing the block is said partial.
Not all block constructions are possible since blocks crossing is not
allowed and the selection of one block may prevent the construction of another
one. For instance, from Fig. \ref{fig:sequences} one can observe that a
block made of factors ``1'' may be created. Another block with factor ``2''
may also be created. However, both blocks cannot be present together in the
alignment.  
These blocks are said \emph{incompatible}. A group of
blocks is said to be \emph{compatible} if all couples of blocks are
compatible. 

Another relation between potential blocks can be observed in their
neighborhood. Indeed, a strong relevance has to be accorded to potential
blocks that are closed to one another and that do not cross. If two factors
are neighbors in many sequences, there is a high probability for these
factors to be part of a bigger factor with little differences. Such relations
are taken into account within our approach and are called friendly relations.
An example of friendly relation in the sample alignment can be observed
sequences 2 and 3 between factors ``1'' and ``2''.     

\begin{figure}
        \centering
                \includegraphics[width=0.9\textwidth]{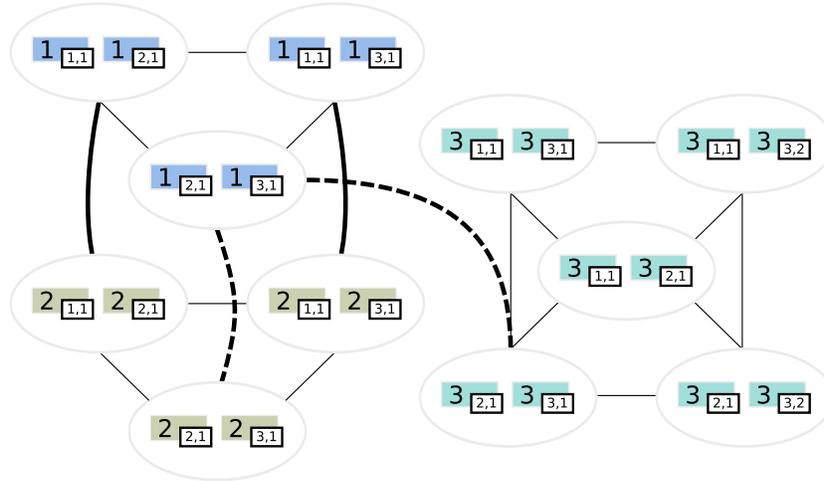}
        \caption[]{The relation graph $G_r$ based on the dual graph of $G$ with
                additional edges corresponding to compatibility constraints,
                represented by thick plain edges, and friendly relations,
                represented by dashed links.} 
        \label{fig:graph2}
\end{figure}

The \emph{factor graphs} are intended to represent the search space of blocks
according to the set of considered factors. 
However, they do not capture compatibility constraints and friendly relations
between blocks. 
For that purpose, we first consider $G'=(V',E')$ dual
graph of $G=(V,E)$, $G$ being the graph composed of the entire set of
\emph{factor graphs}.  
The set $V'$ of $G'$ corresponds to the set $E$ of $G$. Each couple of
adjacent edges $e_1,e_2 \in E$ corresponds to one edge in $E'$. 
Moreover, in order to represent compatibility constraints and friendly relations
two new kinds of edges have to be added to this graph: namely $E_c$ for
compatibility constraints and $E_f$ for friendly relations.
We call \emph{relation graph} (Fig. \ref{fig:graph2}) the graph $G_r= (V',E' \cup E_f \cup~E_c)$.
  
Our approach makes use of both graphs. Ants move within the factor graphs, but
their actions may also produce some effects in remote parts of the factor graphs,
according to relation graph topology as explained in Sect. \ref{sec:algorithm}.

%----------------------------------------------------------------
\subsection{Ants for Multiple Sequence Alignment}
%----------------------------------------------------------------

Ant based approaches have shown their efficiency for single or multiple
criteria optimization problems. The central methodology being known as Ant
Colony Optimization \cite{dorigo97}. Ants in these algorithms usually evolve in a
discrete space modeled by a graph. This graph represents the search space of
the considered problem. Ants collectively build and maintain a
solution. Actually, this construction takes place thanks to pheromone trails laid
down the graph. The search is led both by local information in the graph and
by the global evaluation of the produced solutions. 

The model issued in the previous section proposed a graph model that raises
local conflicts and attractions that may exist in the neighborhood of the
factors in the alignment. For this, the model may handle the local search
needed by classical ACOs. However, providing a global evaluation function for
this problem is unlikely. 
Indeed, defining a relevant evaluation function for MSA is in itself a problem
since the evaluation should be aware of the evolutionary history of the
underlying species, which is part of the MSA problem.
Molecular biologists themselves do not all agree whether
or not one given alignment is a good one. Popular evaluation functions like
the classical \emph{sum of pairs} \cite{Altschul89} are still debated. 
As a consequence, instead of focusing on such a function, our approach
concentrates on building structures in the factor graphs according to the
relation graph. 
These structures correspond to compatible blocks. 
The building of blocks is made by ants which behavior is directly constrained
by the pheromones they laid down in the graph and indirectly by the relation
graph since this graph has a crucial impact on pheromone deposit location. 

The global process can be further refined by taking into account the size of
the selected factors, as well as the number of nucleotides or amino acids
located between the factors of two neighbor blocks. 
These numbers are called \emph{relative distances} between factors in the
sequel. 
Thus, the factor graphs carry local information necessary to the ant system and
acts like the environment for ants. Communication via the environment also
known as stigmergic communication \cite{grasse_59} takes place in
that graph; pheromones trails are laid down on the edges according to ants
move, but also according to the relation graph. 
A solution of the original problem is obtained by listing the set of
compatible blocks that have been bring to the fore by ants and revealed by
pheromones.

\vspace{-0.2cm} 

%-----------------------------------------------------------
\section{Algorithm}
\label{sec:algorithm}
%------------------------------------------------------------

The proposed ant-based system does not evaluate the produced solutions, in
this way it is not an ACO. However, the local search process remains widely
inspired from ACOs. The general scheme of the behavior of the ant based system
follows these rules:
\begin{itemize}
\item Ants perform walks into the factor graphs.
\item During these walks each ant lay constant quantities of phe\-ro\-mones down on the edges they cross.
\item This deposit entails a change in pheromone quantities of some remote edges according to the relation graph $G_r$ as described in Sect. \ref{sec:model}. 
\item Ants are attracted by the phe\-ro\-mone trails already laid down in the environment.

\end{itemize}
Finally a solution to the problem is a set of the most pheromone loaded edges of the \emph{factor graph} that are free from conflicts. 
In the following section pheromone management is more formally detailled.
%What follows illustrates implementation details for this algorithm.

\subsection{Phe\-ro\-mone Trails}

Let $\tau_{ij}$ be the quantity of phe\-ro\-mone present on edge $(i,j)$ of graph $G$ which links nodes $i$ and $j$. If $\tau_{ij}(t)$ is the  quantity of phe\-ro\-mone present on the edge $(i,j)$ at time $t$, then  $\Delta\tau_{ij}$ is the quantity of phe\-ro\-mone to be added to the total quantity on the edge at the current step (time $t+1$). So:
\begin{equation}
 \tau_{ij}(t+1)=(1-\rho).\tau_{ij}(t) + \Delta\tau_{ij}
\end{equation} 
 
Note that the initial amount of pheromone in the graph is close to zero and that $\rho$ represents the  evaporation rate of the phe\-ro\-mones. Indeed, the modeling 
of the evaporation (like natural phe\-ro\-mones) is useful because it makes it 
possible to control the importance of the produced effect. In practice, the 
control of this evaporation makes it possible to limit the risks of premature 
convergence of the process.

The quantity of phe\-ro\-mone $\Delta\tau_{ij}$ added on the edge $(i,j)$ is the 
sum of the phe\-ro\-mone deposited by all the ants crossing the edge $(i,j)$ with 
the new step of time. The volume of phe\-ro\-mone deposited on each passage of an ant is a constant 
value $Q$.
If $m$ ants use the edge $(i,j)$ during the current step, then:

\begin{equation} 
\Delta\tau_{ij} = mQ
\label{eq:deltaTauIJ}
\end{equation}

\subsection{Constraints and Feedback Loops}
Generally speaking, feedback loops rule self-organized systems. Positive feedback loops increase the system tendencies while negative feedback loops prevent the system from 
continually increasing or decreasing to  critical limits. In this case, pheromone trails play the role of positive feedback loop attracting ants that will deposit pheromones on the same paths getting them more desirable. Friendly relationship found in the $G_r$ graph may also play a positive feedback role. Indeed, more pheromones are laid down around friendly linked blocks. On the other side, conflicts between blocks act as negative feedback loops laying down 'negative' quantities of pheromone. 

Let consider an edge $(i,j)$ that has conflict links with some other edges. During the current step, the $c$ ants  that cross edges in conflict with edge $(i,j)$ define the amount of negative pheromones to assign to $(i,j)$: $\Delta\tau_{ij}^{conflict}=cQ$.
 
Besides, an edge $(i,j)$ with some friendly relations will be assigned positive pheromones according to the $f$ ants that cross edges in friendly relation with $(i,j)$ during the current step :   $\Delta\tau_{ij}^{friendly}=fQ$.

Finally, the overall quantity of pheromone on one given edge $(i,j)$ for the current step defined in equation \ref{eq:deltaTauIJ} is modified as follow: 

\begin{equation} 
\Delta\tau_{ij} = \Delta\tau_{ij} + \Delta\tau_{i,j}^{friendly} - \Delta\tau_{i,j}^{conflict}
\end{equation}

\subsection{Transition Rule}
When an ant is on vertex $i$, the choice of the next vertex to be visited must be 
carried out according to a defined probability rule.

According to  the method  classically  proposed  in  ant algorithms, the choice 
of a next vertex to visit is influenced by  2 terms. The first is a local 
heuristic based on local information, namely the \emph{relative distance} 
between the factors ($d$). The second term is representative of the stigmergic 
behavior of the system. It is the quantity of phe\-ro\-mone deposited ($\tau$).

\remark{ Interaction between positive and negative phe\-ro\-mones can lead 
on some edges to an overall negative value of phe\-ro\-mone. Thus, phe\-ro\-mones 
quantities need normalization before the random draw is made. Let  $max$  be an 
upper bound value computed from  the largest quantity of phe\-ro\-mones on the 
neighborhood of the current vertex. The quantity of phe\-ro\-mone $\tau_{ij}$ 
between edge $i$ and $j$ is normalized as $max - \tau_{ij}$.
}

The  function $N(i)$  returns  the list of  vertices adjacent to $i$ (its
neighbors). The next vertex will be chosen in this list. The probability  
for an ant being on vertex $i$, to go on $j$ ($j$ belonging to ($N(i)$) is:

\begin{equation} P(ij)=\frac{[
\frac{1}{max-\tau_{ij}}^\alpha . \frac{1}{d_{ij}}^\beta ]}{\sum_{s \in N(i)} [
\frac{1}{max-\tau_{is}}^\alpha . \frac{1}{d_{is}}^\beta]}
\end{equation}

The parameters $\alpha$ and $\beta$ make it possible to balance the impact of pheromone trails relatively to the \emph{relative distances}.

\vspace{-0.2cm} 

%-------------------------------------------------------------------------------
\section{Analysis}
\label{sec:analysis}

\begin{figure}
        \centering
                \includegraphics[width=1.00\textwidth]{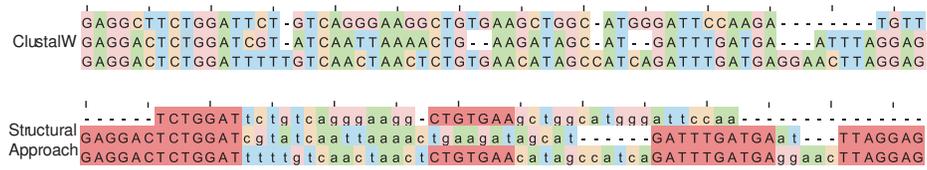}
        \label{fig:comparison}
        \caption{Alignment of 3 sequences of TP53 regulated inhibitor of apoptosis 1 for Homo sapiens, Bos taurus and Mus musculus. Comparison of the alignment given by ClustalW and our structural approach. The red uppercase nucleotides on the structural approach are the blocks.}
\end{figure}

First experiments on real sequences have been
performed. Fig. \ref{fig:comparison} compares some results obtained by the
proposed structural approach and the alignment provided by ClustalW. This sample shows that the visible
aligned blocks can be regained in the results given by ClustalW.  
However, results are still to be evaluated. 
Indeed, it may happen that some blocks are found by our method while they are
not detected by ClustalW.
Anyway, we are aware of the necessity of additional analyses and discussions
with biologists in order to validate this approach. 

\vspace{-0.2cm} 

%-------------------------------------------------------------------------------
\section{Conclusion}
\label{sec:conclusion}
%-------------------------------------------------------------------------------

In this paper was proposed a different approach, for the problem of multiple sequence alignment. 
The key idea was to consider a problem of building and maintaining a structure
in a set of biological sequences instead of considering an optimization problem. 
Preliminary results show that the structures built by our ant-based algorithm
can be informally compared, on a pattern basis, with the results given by
ClustalW.   
 
The outlook for the project is now to prove the efficiency and the relevance
of the method, in particular, an important chunk of future work will concern
the comparison of the differences between conserved regions provided by
ClustalW and other well-known multiple sequence alignment methods and our
approach.   
The second perspective focuses on the performance of the method. Indeed, the
way blocks are built and intermediate results allow us to consider a kind of
divide-and-conquer parallel version of this tool. 
Most recent results and advances will be made available on \texttt{www.litislab.eu}.

\vspace{-0.2cm} 

%-------------------------------------------------------------------------------
\bibliographystyle{plain} 
\bibliography{lssc2007}

\end{document}